\documentclass[fleqn,usenatbib]{mnras}

\usepackage{newtxtext,newtxmath}

\usepackage[T1]{fontenc}

\DeclareRobustCommand{\VAN}[3]{#2}
\let\VANthebibliography\thebibliography
\def\thebibliography{\DeclareRobustCommand{\VAN}[3]{##3}\VANthebibliography}

\usepackage{graphicx}	
\usepackage{amsmath}	
\usepackage[dvipsnames]{xcolor}
\usepackage{footnote}
\makesavenoteenv{tabular}
\usepackage{tablefootnote}
\usepackage{threeparttable}

\newcommand{\siggas}{$\Sigma_{\rm gas}$}	

\newcommand{\qedi}{\citetalias{QEDI}}
\newcommand{\qediii}{\citetalias{QEDIII}}
\newcommand{\fid}{$\Sigma$6.6}
\newcommand{\inngal}{$\Sigma$25}
\newcommand{\supersol}{$\Sigma$6.5-Z2}
\newcommand{\subsol}{$\Sigma$6.5-Z0.2}
\newcommand{\outgal}{$\Sigma$0.8}
\newcommand{\fidh}{$\Sigma6.5$-$h_\mathrm{SN}$}

\newcommand{\aref}[1]{\hyperref[#1]{Appendix~\ref{#1}}}

\defcitealias{QEDI}{QED I}
\defcitealias{Huang24a}{QED II}
\defcitealias{QEDIII}{QED III}
\newcommand{\quokka}{\textsc{quokka}}
\newcommand{\grackle}{\textsc{grackle}}

\title[QED V]{QED V: Variations in metal loading of galactic winds with element nucleosynthetic origin}

\author[Vijayan et al.]{
Aditi Vijayan$^{1}$\thanks{E-mail:aditi.vijayan@anu.edu.au},
Mark R. Krumholz$^{1}$,
Benjamin D. Wibking$^{2}$
\\
$^{1}$Research School of Astronomy and Astrophysics, Australian National University, Canberra ACT 2601, Australia\\
$^{2}$Department of Physics and Astronomy, Michigan State University, 567 Wilson Road, East Lansing, MI 48824, USA\\
}

\date{Accepted XXX. Received YYY; in original form ZZZ}

\pubyear{\the\year{}}

\begin{document}
\label{firstpage}
\pagerange{\pageref{firstpage}--\pageref{lastpage}}
\maketitle

\begin{abstract}
Type Ia supernovae, type II supernovae, and asymptotic giant branch (AGB) stars are important sites of stellar nucleosynthesis, but they differ greatly in their rates, their location within a galaxy, and the mean thermal energy and abundance distribution of their ejecta. In earlier papers in this series we have shown that a significant fraction of metals newly synthesized by type II supernovae are promptly lost to galactic winds -- i.e., galactic winds are metal loaded. Here we investigate whether the elements returned by type Ia supernovae and AGB stars are similarly metal loaded, or whether metal loading varies significantly with nucleosynthetic site. We use a series of high-resolution ``tall box'' simulations of the interstellar medium with the \quokka~GPU-accelerated code, within which we systematically vary the galaxy gas surface density, metallicity, and the scale heights and relative rates of the different nucleosynthetic sources. We show that the metal loadings of galactic winds differ substantially between metals produced by different sources, with typical variations at the level of $\approx 0.3$ dex, a phenomenon we term differential metal loading. Which set of metals suffers preferential loss from this phenomenon varies depending on the galactic environment, and is not easily predictable \textit{a priori}. Our findings call into question the the interpretation of diagnostics of galaxy formation, for example star formation timescales and initial mass functions, based on abundance diagnostics, since the abundance variations upon which these techniques rely are often at levels comparable to those we show can be induced by differential metal loading.
\end{abstract}

\begin{keywords}
galaxies: ISM --- galaxies: starburst --- ISM: jets and outflows --- ISM: structure
\end{keywords}



\section{Introduction}
\label{sec:intro}

It has long been recognised that because different heavy elements are produced by different nucleosynthetic pathways occurring in stars of different masses (see \citealt{Nomoto+2013} for a review), measurements of the relative abundances of those elements in a galaxy can in principle be used to constrain many aspects of that galaxy's history. This is the central insight of galactic chemical evolution (GCE) models, with different abundance ratios used to constrain different aspects of galaxy formation (see \citealt{Maiolino19a} and \citealt{Matteucci21a} for recent reviews). For example, the $\alpha$/Fe ratio is often used as a chemical clock: $\alpha$ elements are primarily produced by short-lived ($\lesssim 40$ Myr) massive stars that end their lives in type II supernovae, while iron peak elements are predominantly produced in smaller stars that undergo type Ia supernovae, with a delay $\sim 1$ Gyr \citep{Maoz17a}, and thus variations the $\alpha$/Fe ratio can be used to deduce the timescale over which a stellar population formed. In elliptical galaxies the $\alpha$/Fe ratio increases with velocity dispersion \citep{Thomas+05}, which has been interpreted as an indication that high velocity dispersion ellipticals form most of their stars in a short, single burst \citep{Conroy+14}. Similarly in the Milky Way, the higher $\alpha$/Fe ratio in the thick disc compared to the thin disc has been taken as evidence that the galaxy formed in two phases, one fast and one slow \citep{Chiappini97a}.  

Nor is the star formation timescale the only application of GCE models. A number of authors have also used $\alpha$/Fe ratios or ratios of $\alpha$ to $s$-process elements produced in asymptotic giant branch (AGB) stars in attempts to constrain the stellar initial mass function (IMF; see \citealt{McWilliam97}, \citealt{Bastian+10}, \citealt{Hopkins2018}, and \citealt{Smith20a}, and references therein). The logic is that the IMF determines the ratio of the numbers of stars formed that will pass through the different available nucleosynthetic channels, and thus altering the IMF then alters ratios of different elements produced. In the case of dwarf spheroidal galaxies, for example, differences in abundance ratios from Milky Way stars have been used to argue for differences in IMF slope or high mass cut-off \citep{Shetrone+01, Venn+04, Tsujimoto11a}. Similar arguments have also been made regarding the Galactic Bulge \citep{Wyse&Gilmore1992} and galaxy clusters \citep{Portinari+04b, Tornatore+04}. And beyond the IMF and star formation timescales, a variety of authors have used chemical abundance ratios as evidence for the importance of other processes in galaxy evolution, for example radial flows of stars and gas \citep[e.g.,][]{Schonrich09a, Minchev13a, Minchev14a}.

However, all of the GCE models that have been used to draw these conclusions rely on an extremely simple treatment of galactic winds that assumes that metal abundances in the outflowing material are identical to those in the interstellar medium (ISM) from which the outflow is launched. There is accumulating evidence from both theory and observations that assumption is likely incorrect. With regard to the observations, direct measurement of the metallicity of galactic winds indicates that they are metal-loaded, meaning that the wind is higher-metallicity than the galaxy from which it emerges \citep[e.g.,][]{Chisholm+2018, Lopez+2020, Lopez+23, Cameron+2021, Hamel-Bravo+24}; indeed, in \citet[hereafter \citetalias{Huang24a}]{Huang24a} we showed that the correlation between wind metallicity and distance from the driving galaxy recently revealed in X-ray observations is naturally explained by partial mixing between a super-metal rich hot phase that carries the bulk of the supernova ejecta out of the galaxy and a comparatively metal-poor cooler phase of entrained ISM.

With regard to theory, there are a number of observations that are difficult to reproduce if one assumes that ISM metallicities and wind metallicities are equal. For example, observations shows that galaxies' circumgalactic media (CGMs) hold between half (for large galaxies) and the great majority (for dwarf galaxies) of their metals, which can be explained without invoking metal-loaded winds only if dwarf galaxies have extreme mass loading factors $\gg 10$ \citep{Tumlinson+11, Peeples&Shankar2011, Peeples+14, Forbes19a, Deepak+25}. Similarly, the relative flatness of galaxy metallicity gradients, and the dependence of their steepness on galaxy mass, is difficult to reproduce without invoking metal-loaded winds \citep{Sharda21b, Sharda24a}. And in a direct confirmation of these indirect inferences, the simulations we have carried out as part of the QED simulation suite have consistently shown that at least some galaxies likely have strongly metal-loaded winds \citep[hereafter \citetalias{QEDI} and \citetalias{QEDIII}, respectively]{QEDI, QEDIII}.

Once one relaxes the assumption that wind abundances must match ISM abundances, it immediately opens up the question of whether the differences between wind and ISM abundances are the same for all elements, or whether winds might be more heavily metal-loaded in some elements than others. If there is substantial variation in metal loading, this might call into question the conclusions about timescales, the IMF, and other aspects of galaxy formation that have been claimed based on abundance ratios in GCE models. Indeed, chemical evolution modellers have sometimes invoked exactly this effect to explain anomalous abundance patterns that would otherwise point to rather extreme variation in the IMF, for example the observed spread in N/O versus O/H and H/He versus O/H in dwarf irregular galaxies \citep{Pilyugin93, Marconi+94}.

However, there has been no systematic numerical study of how metal loading factors vary between elements in a large galaxy like the Milky Way. All of the QED simulations carried out to date have used a single ``metallicity'' field tuned to follow only type II supernova ejecta. While there have been multiple simulations of both isolated galaxies \citep[e.g.,][]{Minchev13a, Minchev14a, Zhang25a} and galaxies in cosmological context \citep[e.g.,][]{Kobayashi11a, Grand17a} that follow multiple elements from different nucleosynthetic sources, such simulations lack the resolution required to study metal loading. Direct prediction of differential loading requires that one resolve the hot-cold phase structure of the galactic wind, and in \citetalias{QEDI} we showed that this requirement is not met, and thus metal loading factors do not converge, until the resolution reaches $\approx 2$ pc. Moreover, this high resolution is required not just in dense regions near the disc, but out to scales of multiple kpc around the galaxy, since it is in these near-disc regions where mixing between the hot, metal-enriched and cool, metal-poor, entrained phases occurs. Such high resolutions have only been achieved in studies of very low-mass dwarf galaxies \citep[e.g.,.][]{Emerick+2018, Emerick19a, Emerick20a, Brauer25a, Mead25b}, not in the more massive galaxies where the bulk of cosmological star formation takes place.

Our work in \citetalias{QEDIII}, however, provides a hint: the main conclusion of our analysis in that paper was that the location of SNe relative to the galactic plane was an important factor in determining the extent of metal loading of outflows. SNe that explode in a rarer medium farther off the plane inject their metals predominantly into the hot phase which promptly escapes the galaxy, leading to poorer metal retention. Since we observe that type II and type Ia supernovae have different vertical distributions \citep{Hakobyan17a}, this strongly hints at the possibility of differential metal loading elements produced via these different channels. And of course AGB stars return elements to the ISM with much lower energy and do not produce a hot phase at all, suggesting that metal return from them may also lead to different amounts of escape from the galaxy compared to processes where metal return is explosive.

This lack of a systematic study combined with the suggestive hints from \citetalias{QEDIII} provide the motivation for this paper. We seek to understand if type II-, type Ia-, and AGB-produced metals are loaded differentially into outflows, and if so to quantify the implications of differential metal loading for elemental abundance ratio-based diagnostics for galaxy formation. The paper is structured as follows: \autoref{sec:methods} details this setup and highlight how it differs from from the previous QED setup. \autoref{sec:results} we discuss the main results for the Solar neighbourhood case, followed by those for different environments. Finally we discuss the implications of our findings in \autoref{sec:discussion}, and summarise and conclude in \autoref{sec:conclusion}.

\section{Methods and Simulations}\label{sec:methods}

In \autoref{ssec:setup} we describe the physical setup of our simulations, and in \autoref{ssec:metals} we describe our method for injecting metals from type II supernovae, type Ia supernovae, and AGB stars. For reader convenience, we summarise the properties of all simulations that we run in \autoref{tab:params}.

\begin{table*}
\begin{center}
\begin{threeparttable}
\begin{tabular}{|l|c|c|c|c|c|c|c|c|c|c|c|}
\hline
\hline
Name & \siggas & $\Gamma_{\rm SN}$ & $h_{\rm II}$ & $h_{\rm Ia}$ & $h_{\rm AGB}$ & $\Gamma_{\rm II}/\Gamma_{\rm SN}$ &  $Z_{\rm bg}$ & $\Delta x$  & $L_z$ &  $t_f$  \\
 &  [M$_\odot$ pc$^{-2}$] &  [kpc$^{-2}$ yr$^{-1}$] & [pc] & [pc] & [pc]
&  & [$Z_{\odot}$] & [pc]  & [kpc] & [Myr] 
 \\
(1)  & (2) & (3) & (4) & (5) & (6)  & (7) & (8) & (9) & (10) & (11) \\
\hline
\hline
$\Sigma$6.5\tnote{$\dagger$} & $6.5$   & $6\times 10^{-5}$   &  $150$ & $300$ & $300$ & $0.6$ & $1$  & $2$ & $4$ & $179$ \\
\\

$\Sigma$-6.5-0.2$Z_{\odot}$\tnote{$\dagger$} & $6.5$   & $6\times 10^{-5}$   &  $150$ & $300$ & $300$ & $0.6$ & $0.2$  & $2$ & $4$ & $359$ \\
\\

$\Sigma$-6.5-2$Z_{\odot}$\tnote{$\dagger$} & $6.5$   & $6\times 10^{-5}$   &  $150$ & $300$ & $300$ & $0.6$ & $2$  & $2$ & $4$ & $98$ \\
\\

$\Sigma$6.5-$h_{\rm SN}$ & $6.5$   & $6\times 10^{-5}$   &  $150$ & $150$ & $150$ & $0.6$ & $1$  & $2$ & $4$ & $165$ \\
\\

$\Sigma$6.5-$\Gamma$ & $6.5$   & $6\times 10^{-5}$   &  $150$ & $300$ & $300$ & $0.9$ & $1$  & $2$ & $4$ & $165$ \\
\\

$\Sigma$25\tnote{$\dagger$} & $25$   & $3.9\times 10^{-4}$   &  $150$ & $300$ & $300$ & $0.6$ & $1$  & $2$ & $4$ & $127$ \\
\\

$\Sigma$0.8\tnote{$\dagger$}& $0.83$   & $1.58\times 10^{-6}$   &  $1000$ & $2000$ & $2000$ & $0.6$ & $1$  & $4$ & $8$ & $434$ \\
\\

\hline
\hline
\\

\end{tabular}
\caption{Summary of parameters for all runs. Column (2) gives the initial gas surface density and (3) is the rate of SN events derived from the Kennicutt-Schmidt relation. Columns (4), (5), (6) tabulate the scale heights of type II supernovae, type Ia superovae, and AGB stars, respectively. Column (7) is the fraction of SNe that are type II SN; the fraction of type Ia is $1- \Gamma_{\rm II}/\Gamma_{\rm SN}$ and the rate of AGBs is fixed at $16\Gamma_{\rm SN}$ for all runs. Columns (8-11) list the other parameters: metallicity ($Z$) which sets the cooling rate of gas, the resolution ($\Delta x$) which is uniform throughout the box, the box half-height ($L_z$), and the total duration for which the simulations have been evolved.}
\label{tab:params}

\begin{tablenotes}
\item[$\dagger$]The equivalent names for these runs in \qediii\ are, from top to bottom, $\Sigma$13-Z1-H150, $\Sigma$13-Z0.2-H150, $\Sigma$13-Z2-H150, $\Sigma$50-Z1-H150, and $\Sigma$2.5-Z1-H1000. See \autoref{fn:error} for an explanation of the change in naming.

\end{tablenotes}
\end{threeparttable}
\end{center}
\end{table*}

\subsection{Simulation setup}
\label{ssec:setup}

Our simulations use the same basic setup as described in \qedi~and \qediii, and we refer readers to those papers for full numerical details. Here we simply summarise the basic setup for reader convenience. Our simulations use the GPU-accelerated \quokka~code \citep{QuokkaMethods, He24a} to solve the Euler equations of gas dynamics for an inviscid fluid that is subject to optically thin radiative cooling. We model cooling using the pre-tabulated cooling rates provided as part of the \grackle~library \citep{Grackle}, which allows cooling to a floor temperature of 10 K.

The simulation domain consists of a box of size $L_x \times L_y \times 2 L_z$, where for all simulations presented in this paper $L_x = L_y = 500$ pc. The domain is periodic in the $x$ and $y$ directions, and has diode boundary conditions at the vertical faces at $z = -L_z$ and $z = +L_z$. The resolution is uniformly high -- 2 pc in all of our simulations but one, which uses 4 pc cells. The gas in the domain is initially at rest, with a uniform density in the $x$ and $y$ directions and a density profile $\rho(z)$ in the $z$ direction chosen so that it is in hydrostatic equilibrium against a static gravitational potential $\phi(z)$ with a minimum at $z = 0$, which represents the gravitational pull of the gas together with the stars and dark matter toward the galactic midplane. The exact method by which we set the potential and density profiles is described in \qediii.


The initial conditions in our simulations are characterised by two parameters: the gas mass per unit area $\Sigma_\mathrm{gas}$ and the mean metallicity $Z$, which affects the cooling rate. We follow the nomenclature introduced in \qediii: runs are identified as $\Sigma XX$-Z$z$, where `$XX$' indicates the initial gas surface density in units of $M_{\odot}$ pc$^{-2}$ and $z$ is the metallicity in units of $Z_{\odot}$; for brevity we omit the $Zz$ portion of the run name for runs at Solar metallicity. Runs in which we have modified some other aspect of the setup are denoted as $\Sigma XX$-AA, where AA describes the modification we have made.  We provide a complete list of all the simulations we run in \autoref{tab:params}. These represent a subset of the cases presented in \qediii.\footnote{Note that, due to a bug in the script that generated the simulation initial conditions, which was only discovered after publication, the gas surface densities used in the \qediii~simulations were slightly different than intended. Thus the names given in the \qediii~paper do not exactly match the true gas surface densities. In this paper we use the same initial conditions as in the \qediii~paper to enable comparison with that work, but we have corrected the names of the runs to reflect the true gas surface densities used. In \autoref{tab:params} we give the corrected run names and gas surface densities, and provide the corresponding names in the \qediii~paper in the notes to the table. \label{fn:error}}

\subsection{Supernova and AGB star metal injection}
\label{ssec:metals}

Compared to \qediii, the main change in this work is that, in addition to type II supernovae, we also include type Ia supernovae and AGB stars.

\subsubsection{Type II supernoave}

Feedback from type II supernovae is implemented identically to our approach in \qediii. Briefly, in any time step where we determine a supernova occurs at a given position, we add $10^{51}$ erg of thermal energy and $\Delta M_\mathrm{SN} = 5$ M$_\odot$ of mass to the cell in which the supernova occurs. We also add a mass $\Delta M_\mathrm{II} = 1$ M$_\odot$ of a passive scalar, enabling us to track the distribution of supernova-injected material as the simulation runs; in what follows, we will denote the density of this passive scalar as $\rho_{Z,\mathrm{II}}$, and we define the metallicity in this scalar normalised to Solar as\footnote{Note that the choice of a single mass $\Delta M_\mathrm{II}$ from all supernovae is obviously a rather crude approximation, but the absolute value of the mass injected will not matter to any of our quantitative results, because we will always normalise out the choice of $\Delta M_\mathrm{II}$ in what follows. Also note that the metallicity $\rho_{Z,\mathrm{II}}$ is distinct from the metallicity $Z$ that sets our cooling rate, which is fixed at the start of the simulations. We do not self-consistently adjust the cooling rate to account for enrichment by supernovae, because doing so would prevent us from doing controlled experiments to determine how different metallicities affect the properties of outflows. See \qediii~for further discussion of this choice.}
\begin{equation}
    \frac{Z_\mathrm{II}}{Z_\odot} = \frac{\rho_{Z,\mathrm{II}}}{\rho Z_\odot},
\end{equation}
where we take $Z_\odot = 0.0086$, the approximate oxygen mass fraction in the Sun \citep{Asplund09}. The rate $\Gamma_\mathrm{SN}$ at which supernovae occur is chosen to match the rates used in the \qediii~simulations, which in turn was calibrated empirically from the Kennicutt-Schmidt relation \citep{Kennicutt98a}; we report the value used in each simulation in \autoref{tab:params}. In the present simulations, we split $\Gamma_\mathrm{SN}$ is split between type II and Ia; for most of our runs type II supernovae account for $60\%$ of the total SN explosions, roughly matching the ratio observed in the present-day Milky Way \citep[e.g.,][]{Krumholz23a}, but we also carry out one simulation in which we change this factor to 90\%, as might be expected for a more actively star-forming galaxy. We report the ratio of the type II supernovae rate to the type Ia rate for each simulation, $\Gamma_\mathrm{II}/\Gamma_\mathrm{SN}$, in \autoref{tab:params}. The positions of type II supernovae are determined randomly, with a uniform distribution i the $x-y$ plane and a Gaussian distribution in the $z-$direction with scale height $h_\mathrm{II}$. The scale heights used in each simulation match those used in the \qediii~simulations, and we refer readers to that paper for a discussion of the choice.

\subsubsection{Type Ia supernovae}

Our treatment of type Ia supernovae is essentially identical to that for type II supernovae. The sole differences that (1) the rate is $\Gamma_\mathrm{Ia} = \Gamma_\mathrm{SN} - \Gamma_\mathrm{II}$, (2) we track type Ia supernova ejecta using a different passive scalar field than for type II's, with each explosion adding a mass $\Delta M_\mathrm{Ia}$ of the type Ia passive scalar, and (3) the scale heights of SNIa and SNII are different. Physically we expect that the progenitor population for type II SNe, being younger, occupies a region closer to the disc, while the type Ia supernovae have a wider vertical distribution. Our choice for most simulations is $h_{\rm Ia}=2\ h_{\rm II}$, roughly the ratio found by observations in nearby galaxies \citep{Hakobyan17a}. However, we again carry out some simulations where we vary this ratio in order to understand its influence. We report the values of $h_{\rm Ia}$ for each simulation in \autoref{tab:params}.

\subsubsection{AGB stars}

Our third source of metals is AGB stars. We treat these like supernovae in that we model them as set of injection events that occur at random points in space and time, modifying the properties of the cell where they occur. However, they differ in how they modify that cell. We assume that each AGB injection adds $\Delta M_\mathrm{ABG} = 1.4$ M$_\odot$ of material, together with an amount of thermal energy corresponding to this injected material having a temperature of $10^4$ K; this is much less than the $10^{51}$ erg added by supernovae. When an AGB injection occurs, we also add a passive scalar, which we track separately from the two passive scalars for type Ia and type II supernovae.

AGB star injection events also differ from supernovae in their rate and vertical distribution. We take the AGB injection rate to be $16\times$ the supernova rate, i.e., $\Gamma_\mathrm{ANB} = 16\Gamma_\mathrm{SN}$; given that there is one type II supernova per $\approx 100$ M$_\odot$ of stars formed, this corresponds to assuming one AGB star per $\approx 10$ M$_\odot$ of stars formed, roughly the correct rate for a \citet{Chabrier2001} IMF. As with type Ia supernovae, AGB stars trace an older stellar population, and thus have a large scale height than type II supernovae. Observations suggest a vertical distribution fairly similar to that of type Ia supernovae \citep[e.g.,][]{Jackson02a}, and thus we take $h_\mathrm{AGB} = h_\mathrm{Ia}$ in all our simulations.

\section{Results}\label{sec:results}

We begin our discussion of the results in \autoref{ssec:fid} using run \fid\ as an example and to establish our analysis framework. We then examine how the results vary with galactic environment using our other runs in \autoref{ssec:environment}.

\subsection{Results for run \fid}
\label{ssec:fid}

\begin{figure*}
    	\includegraphics[width=\textwidth]{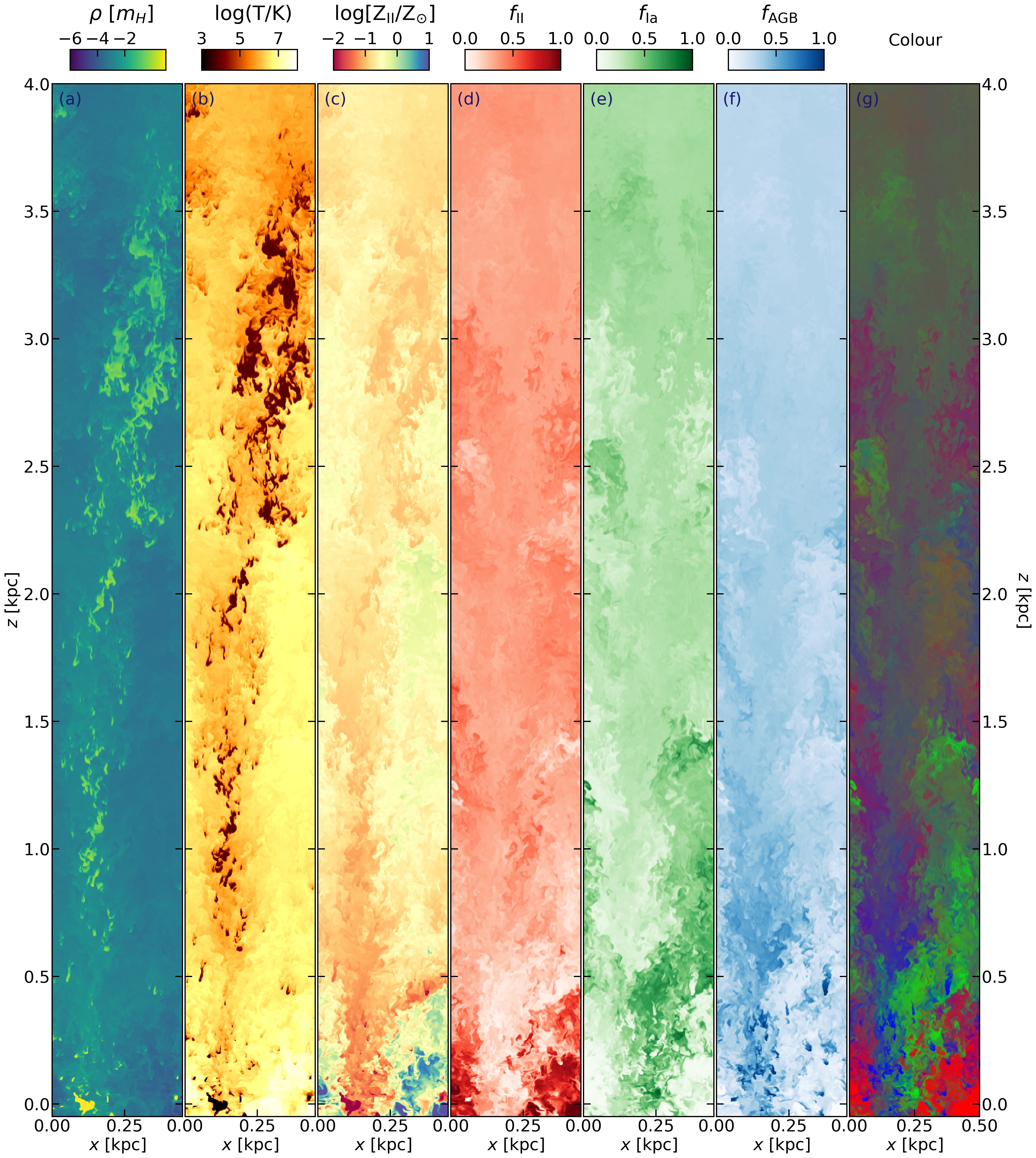}
    \caption{A slice through the plane $y=0$ for run \fid~at $t=60$ Myr. From (a) to (f), the quantities shown are gas density, temperature, SN type II metal abundance normalised to Solar, the metal mixing ratios (\autoref{eqn:colour}) for type II, type Ia, and the AGB. Column (g) shows the colour with the red green, and blue channels mapped to the mixing ratios of type II, type Ia, and AGB ejecta, respectively -- see \autoref{ssec:colour} for details. Note that, for clarity, we show only the upper half of the domain; the full simulation domain extends to $z=-4$ kpc. }
    \label{fig:slice_sig4}
\end{figure*}

In \fid, outflows begin to escape the disc about $10$ Myr after the simulation starts, and the outflow rate becomes nearly constant after $\approx 50$ Myr. To illustrate the morphology of this run, we show a slice through the simulation domain at $y=0$ at $60$ Myr in \autoref{fig:slice_sig4}. Panels (a)-(c) show density, temperature, and the abundance of the type II supernova tracer scaled to the Solar value; (d)-(g) show metal mixing ratio and colour, the definition of which we defer to \autoref{ssec:colour}. As in previous QED simulations, we see a multiphase outflow consisting of cool neutral $< 10^4$ K gas and warm ionised $\sim 10^4$ K gas suspended in a hot ($\gtrsim 10^6$ K) wind. Type II metals are injected in the hot phase and the type II metallicity shown in (c) is therefore strongly correlated with gas temperature and inversely correlated with density. The correlation is strongest near the plane, and weakens with height as mixing between the phases gradually homogenises their abundances.

\subsubsection{Wind metal loading}

\begin{figure}
    	\includegraphics[width=\columnwidth]{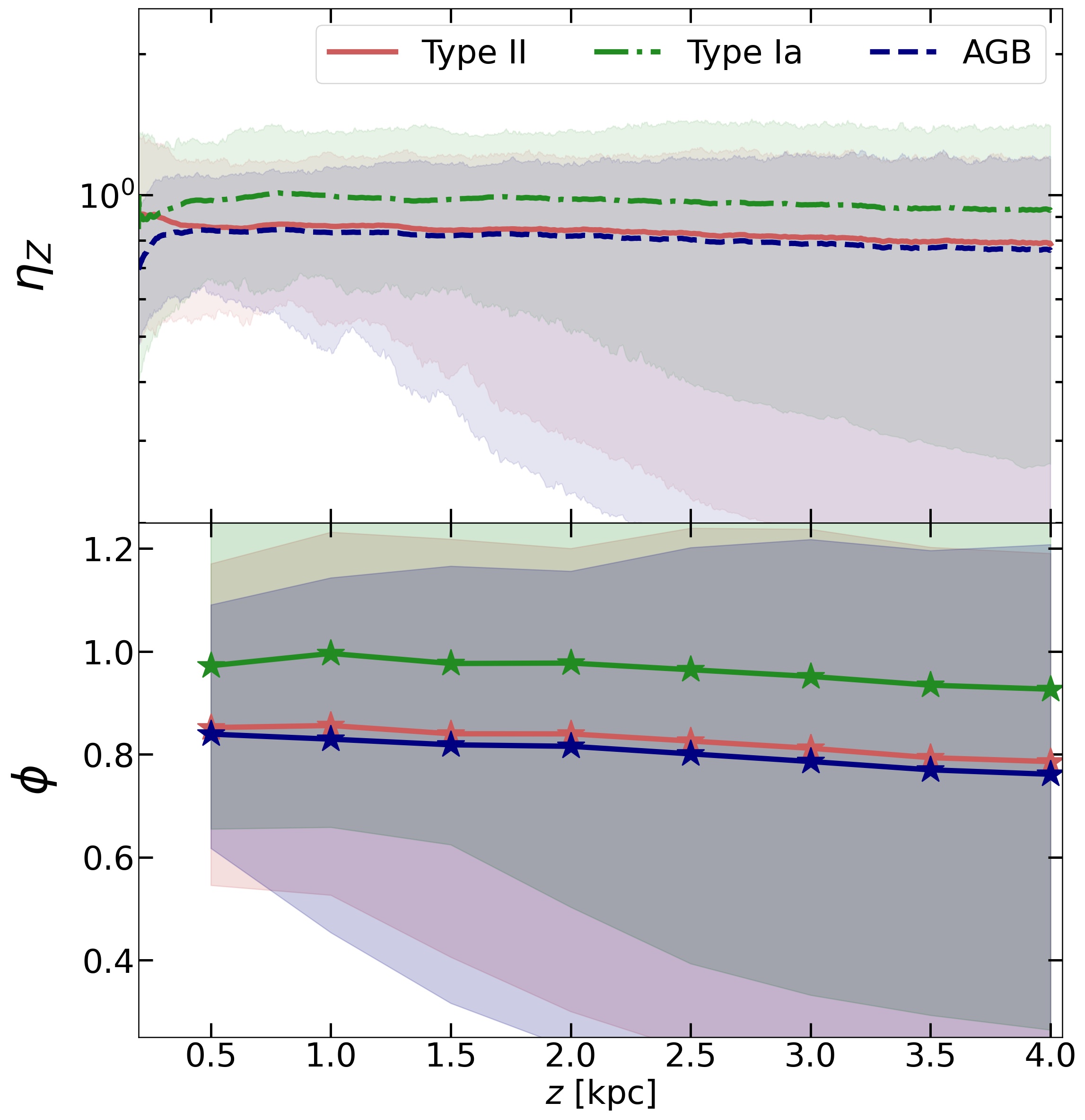}
    \caption{Metal loading factors $\eta_{Z,k}$ (\autoref{eq:etaZ}) and $\phi_k$ (\autoref{eq:phi}) for the \fid\ run. In both panels solid lines and points show time averages, and shaded bands show the $16$th to $84$th percentile variation in time over the full run. Red, green, and blue colours correspond to the metal fields tracing type II ejecta, type Ia ejecta, and AGB ejecta, respectively.}
    \label{fig:outflow_rate}
\end{figure}

Once the simulations reach approximate steady state, we can measure the flux of metals carried outward in the wind. We define the mass flux of each passive scalar through height $z$ at each time as
\begin{equation}
    \dot{M}_{Z,k} = \int_{-L_y/2}^{L_y/2} \int_{-L_x/2}^{L_x/2} \left[\rho_{Z,k} v_z (z) - \rho_{Z,k} v_z (-z)\right] \ dx \ dy,
\end{equation}
where $k$ is II, Ia, or AGB, for the three distinct passive scalars and $v_z$ is the $z$ component of the velocity. Thus our mass flux is a sum of the outward fluxes through the $+z$ and $-z$ surfaces.

We can use this definition of $\dot{M}_Z$ to construct the two metal loading factors of interest, $\eta_{Z,k}$ and $\phi_k$, for each metal type. The former is defined by
\begin{equation}
    \eta_{Z,k} = \frac{\dot{M}_{Z,k}}{\Gamma_{k}\Delta M_k},
    \label{eq:etaZ}
\end{equation}
and can be understood as simply the ratio of the instantaneous metal outflow rate to the instantaneous metal injection rate. The latter is defined as
\begin{equation}
    \phi_k = \frac{\dot{M}_{Z,k} - \langle Z_k\rangle\dot{M}}{\Gamma_k \Delta M_k},
    \label{eq:phi}
\end{equation}
where
\begin{equation}
    \dot{M} = \int_{-L_y/2}^{L_y/2} \int_{-L_x/2}^{L_x/2} \left[\rho v_z (z) - \rho v_z (-z)\right] \ dx \ dy,
\end{equation}
is the total mass outflow rate and
\begin{equation}
    \langle Z_k\rangle = \frac{\int_{-z}^{z} \int_{-L_y/2}^{L_y/2} \int_{-L_x/2}^{L_x/2} \rho_{Z,k}\, dx\,dy\,dz}{\int_{-z}^{z} \int_{-L_y/2}^{L_y/2} \int_{-L_x/2}^{L_x/2} \rho\, dx\,dy\,dz}
\end{equation}
is the mean metallicity of material within a distance $z$ of the midplane for each metal type. As discussed in \qediii, we can intuitively understand $\phi$ as follows: the numerator is the difference between the measured metal flux $\dot{M}_{Z,k}$ and the metal flux we would expect if the outflow consisted of uniformly-mixed material, $\langle Z_k\rangle\dot{M}$. Thus the numerator is the \textit{excess} of metal in the outflow arising from the fact that it contains a mixture of entrained ambient ISM and unmixed stellar ejecta, and we can therefore think of $\phi_k$ as a corrected version of $\eta_{Z_k}$ that accounts for the fact that, in an ISM with finite background metallicity, a wind would contain a finite metal flux even if it were driven by processes that themselves injected no metals. This is the crucial quantity for galactic chemical evolution, since $\phi$ therefore indicates the fraction of newly-synthesised metals that are lost to a galaxy, while $1-\phi$ is the fraction of metals retained.

We plot the time-averaged value and time-variation of $\eta_{Z,k}$ and $\phi_k$ as a function of height $z$ for all three metal types in the upper and lower panels of \autoref{fig:outflow_rate}. That these quantities are relatively close to unity for all three means that the outflows for all types are heavily metal-loaded. However, we can also see that the metal loading differs substantially between the material types, with type Ia ejecta more loaded than the other two. The difference is means is very physically significant -- recall that $1-\phi$ is the fraction retained by the galaxy, so \autoref{fig:outflow_rate} shows that galaxies will lose $\approx 90\%$ of their type Ia ejecta promptly to winds and retain only $\approx 10\%$, while for type II and AGB ejecta they will lose $\approx 80\%$ and retain $\approx 20\%$, leading to factor of two differences in the effective rates of metal enrichment. We discuss the implications of this finding further in \autoref{sec:discussion}.

\subsubsection{Ejecta distribution and mixing}
\label{ssec:colour}

\begin{figure}
    	\includegraphics[width=\columnwidth]{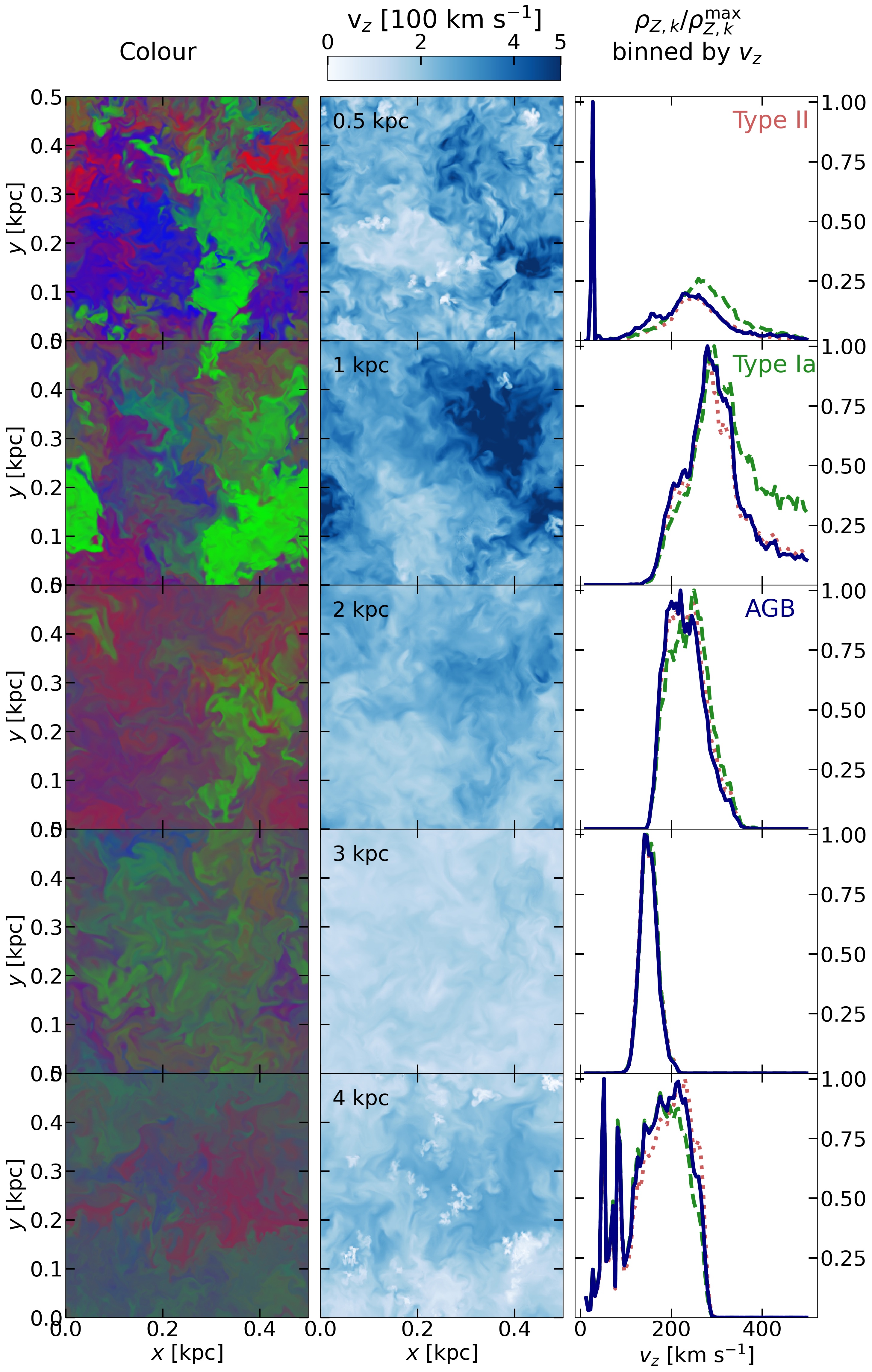}
    \caption{Slices through run \fid~at the same time as shown in \autoref{fig:slice_sig4}. Each row shows a slice in the $xy$ plane at a height $z$ indicated in the legend. The left column shows colour (as in the right panel of \autoref{fig:slice_sig4} -- see \autoref{ssec:colour}), while the middle column shows vertical velocity $v_z$. The right column shows the distribution of metal mass with respect to velocity for each of the three metals -- $Z_\mathrm{II}$ (red dot-dotted), $Z_{\rm Ia}$ (green dashed), and $Z_\mathrm{AGB}$ (blue solid) -- in the slice.}
    \label{fig:colour_height}
\end{figure}

\begin{figure*}
    	\includegraphics[width=\textwidth]{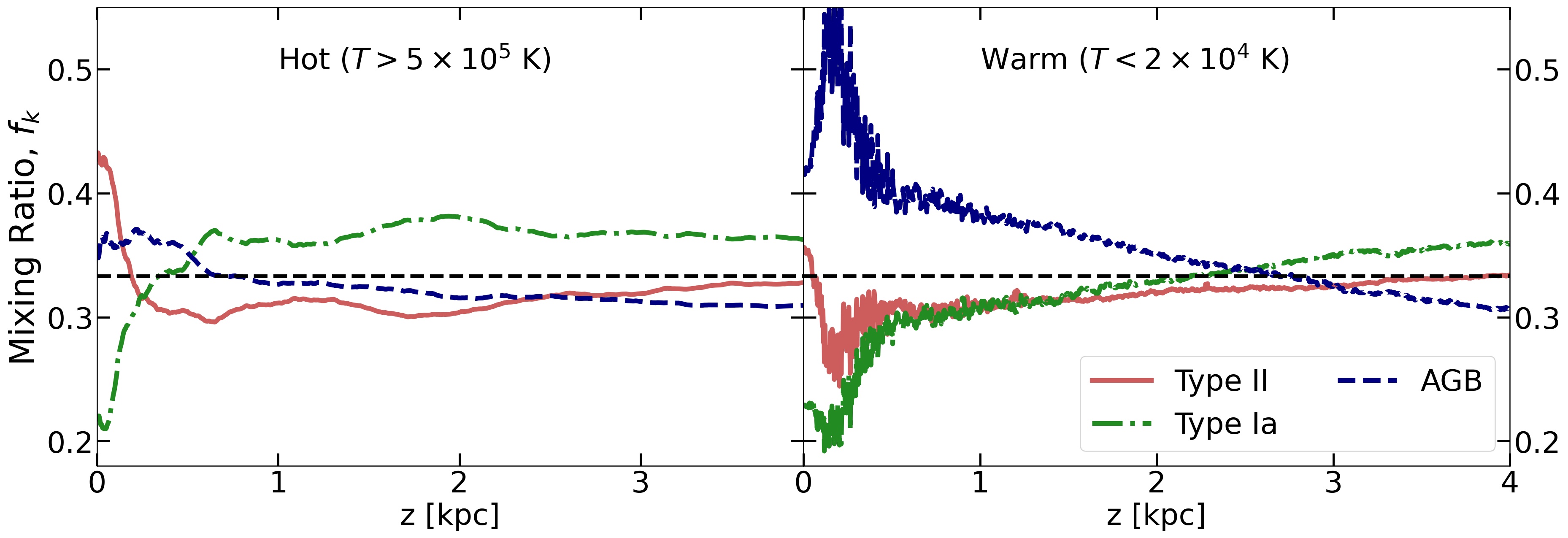}
    \caption{Time-averaged mixing ratio (\autoref{eqn:colour}) as a function of height for gas that is hot ($T > 5\times 10^5$ K; left) and warm ($T < 2\times 10^4$ K; right) for type II (red), type Ia (green), and AGB (blue) ejecta. The quantities plotted are averages over all cells and over all times $t>50$ Myr, the approximate time at which the outflow reaches steady-state. Black dashed horizontal lines indicate $1/3$, the value corresponding to material that is fully mixed.}
    \label{fig:colour_phase}
\end{figure*}

To better understand why the metal loading is different for different types of metals, we define the metal mixing ratio for each metal type $k$ and each position as
\begin{equation}\label{eqn:colour}
    f_k = \frac{(\Gamma_k \Delta M_k)^{-1}\rho_{Z,k}}{\sum_{i} (\Gamma_i\Delta M_i)^{-1}\rho_{Z,i}}.
\end{equation}
We can understand the meaning of this quantity as follows: $\rho_{Z,k}$ is the metal density at a given position (and time), and $\Gamma_k \Delta M_k$ is the rate at which mass of that metal is added, so the numerator represents the metal density at a given point in space and time normalised by the total rate at which that metal is injected. The denominator is simply the sum of this over all three metal fields we follow. Thus if we were to integrate the numerator and denominator over the full simulation volume at fixed time, and if no metals were lost from the simulation domain, the ratio that defines $f_k$ would be exactly $1/3$ for all metal types at all times. Deviations from this ratio indicate must be due either to preferential loss of one of the metal types from the domain, local variations in the abundances of the different metal types, or both. By construction we can immediately see that $\sum_k f_k = 1$, and the possible range of $f_k$ for each metal type runs from 0 to 1, with 0 indicating none of that metal is present and 1 indicating that only that metal is present. We show the value of $f_k$ for each of our three metal types in our example slice in columns (d) - (f) of \autoref{fig:slice_sig4}.

In column (g) of \autoref{fig:slice_sig4} and the left column of \autoref{fig:colour_height}, we show the ``colour'' derived from $f_k$ for each metal type for the same snapshot as that shown in the other columns of \autoref{fig:slice_sig4}. Column (g) of \autoref{fig:slice_sig4} shows the same vertical slice through the simulation box as the remaining panels, while \autoref{fig:colour_height} shows horizontal slices at heights of 0.5, 1, 2, 3, and 4 kpc. The rgb colour scale in these plots is derived from \autoref{eqn:colour}, with the red, green, and blue channel intensities (on an intensity scale from 0 to 1) set to $f_\mathrm{II}$, $f_\mathrm{Ia}$, and $f_\mathrm{AGB}$, respectively. Thus uniformly-mixed material in this plot appears as grey, corresponding to rgb value $(1/3,1/3,1/3)$, while material that is dominated by one of the metal types will appears as bright red, green, or blue. 

At the bottom of \autoref{fig:slice_sig4}, we see a preponderance of red, representing type II ejecta, because the other two sources are injected at higher altitudes, but green and blue become dominant at $\approx 0.5$ kpc, as is clear both from \autoref{fig:slice_sig4} and from the top row of \autoref{fig:colour_height}. This height corresponds roughly to the peak of where type Ia and AGB injection occur. More generally we see from both figures that at lower heights the colours are well separated spatially because the different ejecta types have not yet had the time to mix, and that the metal abundances are at least somewhat correlated with gas velocities (middle and right columns of \autoref{fig:colour_height}) and with gas density. By contrast, beyond 3-3.5 kpc the metals in the outflows are very well mixed, and the outflow velocity has become rather uniform as can be seen in the bottom right panel of \autoref{fig:colour_height}. However, it is clear that even in this well-mixed region the colour is slightly tilted to green, reflecting the preferential loss of type Ia ejecta we saw in \autoref{fig:outflow_rate}.

The middle column of \autoref{fig:colour_height}, which shows the vertical velocity $v_z$ in the same horizontal slices as the mixing ratios shown in the left column, provides some insight as to how this imbalance arises. Examining the first few rows, which show $z=0.5-2$  kpc, it is clear that higher velocities are systematically correlated with regions dominated by type Ia ejecta (indicated by green colour), while regions dominated by AGB ejecta (blue) or type II ejecta (red) have systematically lower velocities. The higher velocities of type Ia ejecta at small height translate directly to higher abundances of these ejecta at large height. We quantify this correlation in the right column of \autoref{fig:colour_height}, which shows the probability density distribution of $\rho_{Z,k}$ for each metal $k$ with respect the outflow velocity $v_z$, with red, green, and blue line colours corresponding to the same metal types -- type II, type Ia, and AGB, respectively -- as in the left panel. At lower heights ($z=0.5, 1.0, 2$ kpc), we see systematic difference between this distributions, with type Ia biased towards higher velocity. As the metals mix their velocity structure becomes uniform, as is visible in the bottom panel at $z=4$ kpc.



The higher velocities of type Ia ejecta in turn are directly related to the characteristic temperature of the gas in which they find themselves, which we illustrate in \autoref{fig:colour_phase}. In this figure, lines show the mean mixing ratio as a function of height for material that is hot ($T > 5\times 10^5$ K; left panel) and warm ($T < 2 \times 10^4$ K; right panel). We first note that type Ia dominates the hot phase for all heights beyond $0.5$ kpc, while type II SNe and AGB contributions are fairly similar. This is merely reflects that measurably more type Ia metals are launched into the hot phase compared to the other two. The warm phase has a different blend of the metals. AGB naturally dominates at lower heights because all its metals are injected in this phase. As the different phases mix warm phase is enriched with type Ia and type II metals and the mixing ratios converge toward the value of $1/3$ that denotes equality. Thus the combination of phase and velocity information yields a consistent picture: more type Ia ejecta are lost because they are preferentially deposited in hotter, faster-moving material that carries them upward and out of the galaxy. This imbalance between the phases is established fairly close to the galactic plane, $z \sim 0.5$ kpc, where type Ia injection primarily occurs.

\subsection{Variation of metal loading with environment}
\label{ssec:environment}

Having discovered and understood differential metal loading for Solar neighbourhood-like conditions, we now proceed to examine the remainder of our runs, which explore different conditions.

\subsubsection{How much does metal loading vary?}

\begin{figure}
    	\includegraphics[width=\columnwidth]{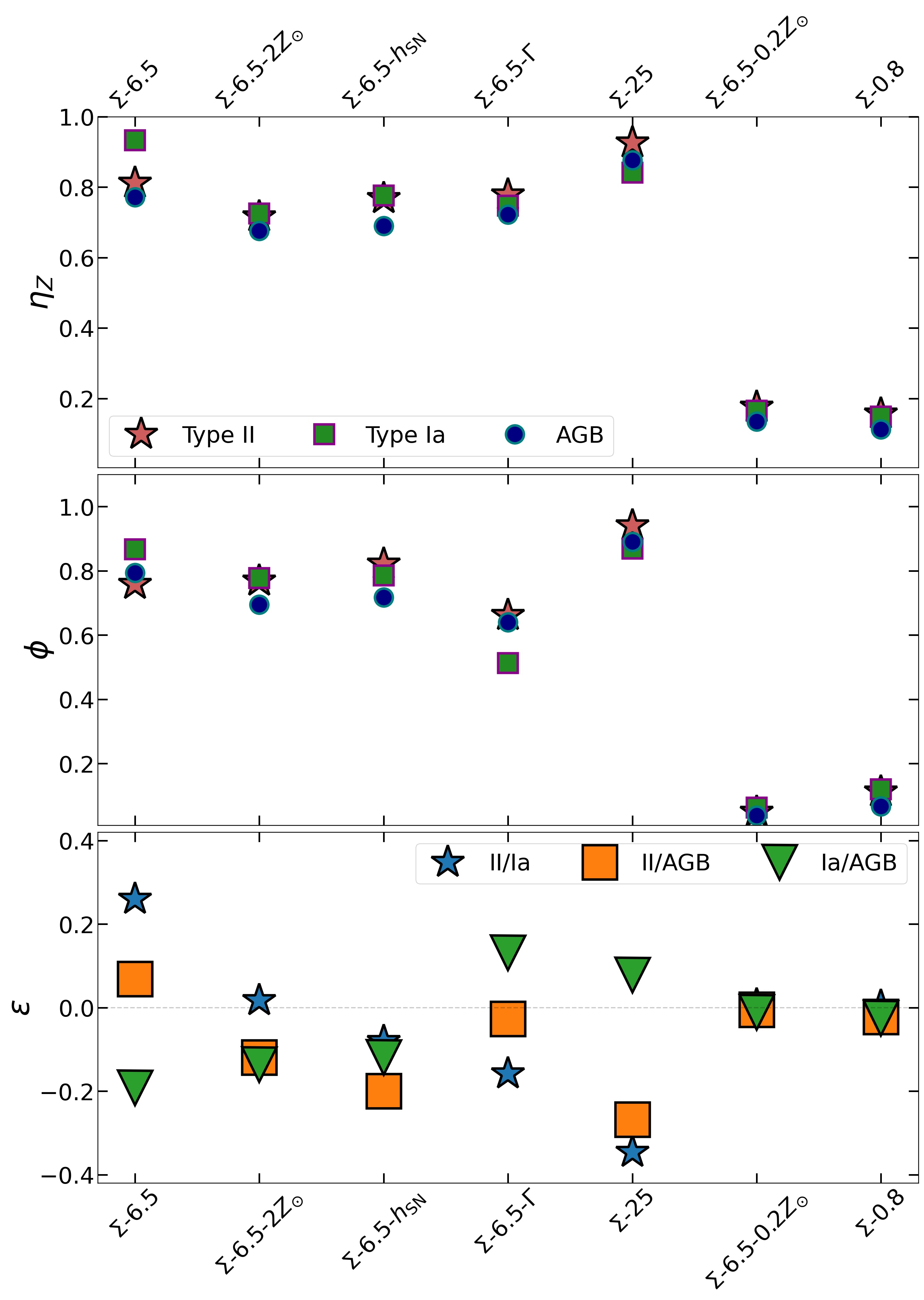}
    \caption{Time-averaged metal loading factors $\eta_Z$ (top; \autoref{eq:etaZ}) and $\phi$ (middle; \autoref{eq:phi}), and differential metal retention $\epsilon$ (bottom; \autoref{eq:epsilon}) for all metal types in all simulations.}
    \label{fig:metal_loading}
\end{figure}

We now repeat the analysis presented in \autoref{ssec:fid} for all the runs listed in \autoref{tab:params}. For each run, we measure the time-averaged value of the two metal loading factors $\eta_Z$ and $\phi$, and summarise these factors in \autoref{fig:metal_loading}. In addition to these two quantities, shown in the top two panels, in the bottom panel we also plot the quantity
\begin{equation}
    \epsilon_{ij} = \mathrm{log} \left ( \frac{1-\phi_{i}}{1 - \phi_{j}} \right),
    \label{eq:epsilon}
\end{equation}
where as usual $i$ and $j$ are II, Ia, and AGB, corresponding to the three metal fields we follow. To understand the physical meaning of this quantity, recall that $\phi_i$ is the fraction of a given metal that is lost promptly to the wind, so $1-\phi$ is the fraction retained in the galaxy; thus the yield of any given metal is effectively reduced by a factor $1-\phi$. Since the ratio of yields in turn determines the abundance ratio for any given element, we can understand $\epsilon_{ij}$ as quantifying the amount by which we would expect differential metal loading to alter the abundance ratio, expressed in units of dex. Thus for example if stellar nucleosynthesis alone would produce abundances of two elements $i$ and $j$ that differ 1 dex, but $\epsilon_{ij} = -0.3$, then we would expect to observe an abundance ratio of only 0.7 dex in the ISM, due to preferential loss of element $i$ relative to element $j$ into the CGM.\footnote{We emphasise however that this might be short-term (compared to the Hubble time) effect, since some of the material that is lost might return on timescales longer than those for which we run our simulation, but still much smaller than a Hubble time. We are unable to address the question of element return on timescales $\gtrsim 0.5$ Gyr given our current simulations.}

From the figure, it is clear that systems that host sustained hot or multiphase outflows, \fid, \inngal\ and \supersol, exhibit large values of $\eta_Z$ and $\phi$, while those that host cool and bursty outflows, \subsol\ and \outgal, have small metal loading and consequently no significant differential metal loading, $\epsilon\approx 0$ for all metals. For the cases with large metal loading, by contrast, there are measurable differences in the values of $\eta_Z$ and $\phi$, which translate to significant differences in $\epsilon$ from run to run. For the case \fid~that we explored in detail in \autoref{ssec:fid} (leftmost column in \autoref{fig:metal_loading}), we see that type Ia elements have the highest metal loading $\phi$, and this translates to $\epsilon_\mathrm{II/Ia} \approx 0.3$ and $\epsilon_\mathrm{Ia/AGB} \approx -0.2$, meaning that differential metal loss is expected to reduce the abundance of type Ia-produced elements in the ISM by $\approx 0.3$ dex compared to type II-produced ones, and by $\approx 0.2$ dex compared to AGB ones, relative to what we would expect based on nucleosynthesis alone. However, not all simulations show the same pattern: for example, in \supersol~(second column) the difference between elements is significantly smaller than in \fid, and $\epsilon_{ij}$ values are closer to zero, indicating little effect of differential metal loading. For \inngal, it is type II rather than type Ia ejecta that are preferentially lost, leading to $\epsilon_\mathrm{II/Ia} \approx \epsilon_\mathrm{II/AGB} \approx -0.3$. And in run \fidh, where have have deliberately made the scale heights of all metal sources equal, we find that type II supernova ejecta are strongly preferentially lost compared to AGB ejecta ($\epsilon_\mathrm{II/AGB} \approx -0.2$) and mildly lost compared to type Ia ejecta ($\epsilon_\mathrm{II/Ia} \approx -0.1$). Clearly there is a non-trivial relationship between scale height and differential metal loss.

\subsubsection{What controls differential metal loss for supernovae?}

\begin{figure*}
    	\includegraphics[width=\textwidth]{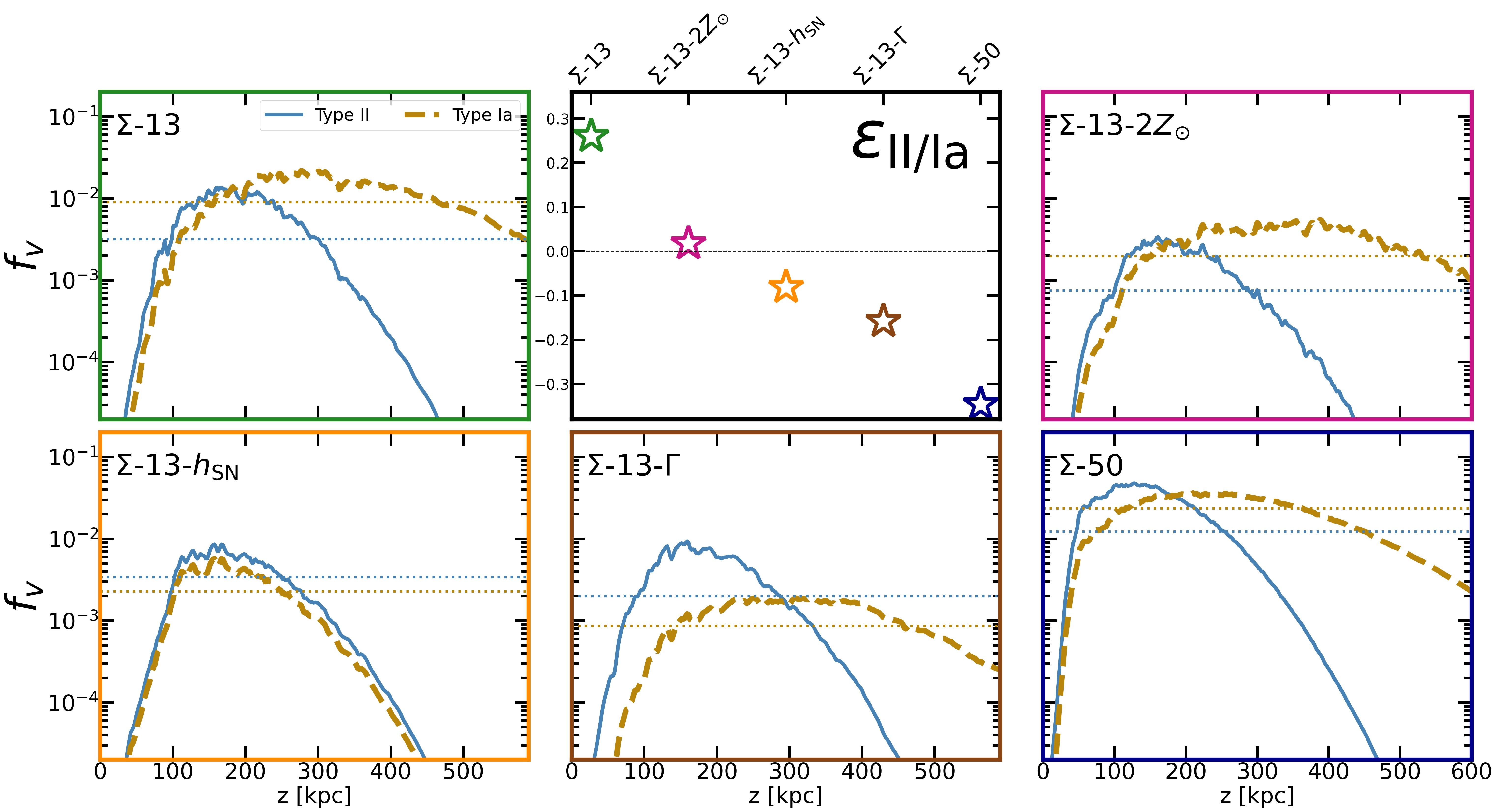}
    \caption{The coloured panels around the edge show the volume filling factor $f_V$ (\autoref{eqn:f_V}), for type II (blue solid) and type Ia (orange dashed) SNe as a function of height, time-averaged over all simulation times after initial breakout of the SNe from the disk (see main text). The horizontal dashed lines indicate the averages over the respective scale heights of the two SN types. In the black panel (top centre), we re-plot $\epsilon_\mathrm{II/Ia}$ from \autoref{fig:metal_loading}, with the colour of the star matching the colour of the corresponding panel showing $f_V(z)$ for that simulation. These panels are ordered from largest to smallest $\epsilon_\mathrm{II/Ia}$ starting from the top left and proceeding first across the top row, then across the bottom row.
    }
    \label{fig:filling_factor}
\end{figure*}

We have seen that differential metal loss occurs in most but not all of the simulations with large metal loading, but that which elements are preferentially loaded varies significantly between runs. Examining \autoref{fig:colour_phase}, and the analogous plots for other simulations (not shown), it is also clear that differential metal loss is established close to the galactic plane, and that beyond $\approx 1$ kpc from the plane the mixing ratio in the hot phase, which carries most of the metal flux, is nearly invariant. Thus the question becomes: what physical processes in the near-disc region determine whether a particular galaxy will lose more of one element type or another?

To try to answer this question, we examine the environments in which type Ia versus type II supernovae occur in each simulation. This is of interest because local environment can play a crucial role in development of a supernova remnant. Density enhancements can retard the expansion of a remnant while also increasing cooling losses, so SNe that explode in a denser environment can experience significant losses on their way out of the disc. Collectively, if a class of SNe tend to occur in a denser environment, this may prevent them from produce a volume-filling hot phase that is capable of breaking out of the disc -- as happens in runs $\Sigma13$-Z0.2 and $\Sigma2.5$, where almost no hot gas escapes. Conversely, if SN bubbles are able to overlap with one another and produce a volume-filling hot phase, they
are more likely to retain their momentum and metal content and 
suffer lesser cooling losses.

To quantify this effect, first consider a SN remnant (SNR) that has developed to the snowplough stage and is consequently slowing as it expands and sweeps up more material. The SNR will begin to fade into and mix with the ambient medium once
the shock speed falls to a value comparable to the sound speed in the medium. The time and SNR radius at which this occurs are \citep{Draine}
\begin{eqnarray}
    t_{\rm fade} & = & 1.87 \, E_{51}^{0.32}  n_0 ^{-0.37}  c_{s,1}^{-7/5} \, \mathrm{Myr}\\
    R_{\rm fade} & = & 67 \, E_{51}^{0.32} n_0^{-0.37} c_{s,1}^{-2/5}\, \mathrm{pc},
\end{eqnarray}
where $E_\mathrm{51}$ is the SN energy in units of $10^{51}$ erg, $n_0$ is the ambient number density in units of H nuclei cm$^{-3}$, and $c_{s,1}$ is the ambient sound speed in units of 10 km s$^{-1}$. We can use these expressions to define an approximate filling factor for SNRs in a region with a specified SN rate per unit volume $\gamma_\mathrm{SN}$, which in the context of our simulations is
\begin{equation}
    \gamma_\mathrm{SN} = \frac{\Gamma_\mathrm{SN}}{\sqrt{\pi} h_\mathrm{SN}} e^{-z^2/h_\mathrm{SN}^2}.
\end{equation}
This is \citep{Draine}
\begin{equation}
    f_V = 1 - \exp\left(-\frac{4\pi}{3} R_\mathrm{fade}^3 t_\mathrm{fade} \gamma_\mathrm{SN} \right).
    \label{eqn:f_V}
\end{equation}
Physically $f_V$ gives the probability that a SN will go off within the ``fadeaway'' volume of another SN under the simple approximation (which is true for our simulation) that SN locations are uncorrelated.

To test our hypothesis that SN overlap is an important factor in metal escape, we can compute the volume averages of density and temperature in every horizontal plane, and use this together with the rate and scale height for both type Ia and type II SNe to obtain $f_V(z)$ for both types in all simulations. For simplicity we do so independently for each SN type, neglecting potential overlap between SNRs produced by SNe of different types. In \autoref{fig:filling_factor} we show the resulting $f_V(z)$s for different runs; the lines shown are time-averages over times $t > t_\mathrm{break}$, where $t_\mathrm{break}$ is the time at which the outflow first breaks out of the disk -- we take this to be $t_\mathrm{break}\approx 50$ Myr for all simulations except \inngal, for which we take $t_\mathrm{break}\approx 20$ Myr due to the more rapid breakout that occurs when the disk is thinner. The blue and orange lines in the figure show filling fractions for type II and Ia's, respectively, and the corresponding horizontal line represents the average over one SN scale height for each type\footnote{We have verified that the results do not change qualitatively if we average over the gas scale height rather than the supernova scale height.}; we refer to this quantity as $\langle f_{\mathrm{V},k}\rangle$.

The figure shows that is a definite correlation between the average filling fraction and metal retention. Run \fid\ has the largest value of $\epsilon_\mathrm{II/Ia}$ (meaning that a higher proportion of type Ia ejecta are lost than type II ejecta), and also the largest value of $\langle f_\mathrm{V,Ia}\rangle - \langle f_\mathrm{V,II}\rangle$ (meaning that type Ia SNe are more volume-filling). As the latter quantity decreases for the other runs, so does $\epsilon_\mathrm{II/Ia}$ -- i.e., \supersol\ has the next-largest value of both $\langle f_\mathrm{V,Ia}\rangle - \langle f_\mathrm{V,II}\rangle$ and $\epsilon_\mathrm{II/Ia}$, followed by $\Sigma$-13-$h_{\rm SN}$ and $\Sigma$-13-$\Gamma$.
 
The one exception to this pattern is \inngal, which has a value of $\langle f_\mathrm{V,Ia}\rangle - \langle f_\mathrm{V,II}\rangle$ similar to that of \supersol, but the smallest value of $\epsilon_\mathrm{II/Ia}$ of any run, indicating preferential loss of type II metals.  A key difference in the filling factor for \inngal, compared to the other runs, lies in the shape of the filling factor curve: for \inngal, the filling factor for both type Ia and type II SNe climbs sharply with height at $z\lesssim50$ pc, a consequence of the small gas scale height in this run, so at $\sim 100$ pc, the filling factor for type II metals is about an order of magnitude larger for \inngal\ compared to \fid. Thus if we were to average over a smaller height for \inngal, it would follow the trend shown by the other simulations.

Thus a tentative but plausible hypothesis we can draw from our simulations is that the relative escape of type II versus type Ia ejecta is the interaction between the relative scale heights of the SNe and the gas, which in turn determines how volume-filling the SNe will be. Whichever SN type has larger volume filling factor over the region where most of those SNe explode will more efficiently ejecta metals from the galaxy.

\section{Discussion}
\label{sec:discussion}

In this section we first in \autoref{ssec:comparison} situate our results within the context of other numerical studies of metal-loaded galactic winds, and then in \autoref{ssec:implications} work through some of the implications of our results.

\subsection{Comparison with other simulation work}
\label{ssec:comparison}

Though there are a very large number of published simulations exploring the properties of galactic winds \citep[and references therein]{Thompson2024}, very few have focussed on the differential loading of elements. In large part this is because such questions are inaccessible at the resolutions typically possible in even zoom-in cosmological simulations aimed at studying galaxy metal abundances \citep[e.g.,][]{Grand17a}. In such simulations typical mass resolutions are $\sim 10^3$ M$_\odot$ at best, and thus different parts of the IMF that give rise to different nucleosynthetic channels cannot be separated. Indeed, even at the $\sim 100$ M$_\odot$ resolutions available in isolated-galaxy simulations following multiple nucleosynthetic channels \citep[e.g.][]{Zhang25a}, metal loading is difficult to study. Higher-resolution simulations that can separate yields from different channels are generally limited to cosmological simulations that explore only the extremely early stages of galaxy formation, or to non-cosmological simulations of isolated galaxies.

In the former category, the \textsc{Aeos} simulations \citep{Brauer25a, Mead25b} follow metal return from multiple nucleosynthetic channels -- in particular focusing on supernovae from population III stars in very early galaxies at $z\approx 15$ with $\sim$pc-scale resolution. They find that the assumption of homogenous mixing of the ejecta is generally invalid, and this inhomogeneities are likely to contribute significantly to the abundances spreads seen in surviving population II stars today. This conclusion is therefore qualitatively consistent with our finding that metals from different nucleosynthetic sources do not mix homogenously, and that this inhomogeneity can leave imprints on chemical abundance ratios.

In the latter category, several authors have simulated outflows from isolated dwarf galaxies including multiple nucleosynthetic sources. In early work, \citet{Recchi+01} present results from a 2D axisymmetric HD simulations of gas-rich dwarf galaxies including both type Ia and type II SNe. They find that the outflows are differentially loaded, with type Ia metals lost more readily than type II metals. They argue that this is because type Ias are injected in hotter gas, consistent with our hypothesis that filling factor determines differential metal loading, since injection in regions of higher temperature and lower density will lead to higher filling factor. More recently, \citet{Emerick+2018, Emerick19a, Emerick20a} simulated outflows with metal loading from AGB star winds, neutron star mergers, SNe, and hypernovae. They find that the metal loading depends largely on the energy of the metal deposition channel, with only $\approx 60\%$ metal loss for AGB stars up to $\approx 95\%$ loss for hypernovae. This is somewhat in contrast to the trends we find, where AGB metals do show lower metal loading than supernovae in some of our simulations, but clearly not all of them, and there is no consistent trend. An important difference in our approaches that may contribute to this is that we place AGB stars systematically farther off the disc than type II supernovae, consistent with observations of the scale height distribution, whereas \citet{Emerick20a} assign identical spatial distributions to the different injection types. However, given the very large differences in other aspects of the simulations -- \citeauthor{Emerick20a} simulate a very small dwarf galaxy in comparison to our simulations parts of a much larger galaxy -- it is difficult to identify a single cause for this difference.

\subsection{Implications for abundance ratio diagnostics}
\label{ssec:implications}

The central finding from our simulations is that differential metal loss can induce differences between true and wind-adjusted nucleosynthetic yields (i.e., yields adjusted downward to account for metals lost promptly to the galactic wind) of up to $\approx 0.3$ dex. These variations are not necessarily in a predicable direction, and depend on galactic environment. In the absence of a full theoretical accounting for how differential metal loss depends on environment, something that our current suite of simulations is not yet broad enough to produce, we must regard the possibility of differential metal loss as a systematic uncertainty in all abundance ratio diagnostics. It is therefore of interest to ask about the extent to which this uncertainty potentially undermines conclusions based on these diagnostics.

In at least some cases the answer appears to be yes. Referring back to some of the examples given in \autoref{sec:intro}, \citet{Conroy+14} find in their sample of early-type galaxies (ETGs) that the abundance ratios of several light elements (C, N, O, Mg Si, and Ti) to iron increase systematically from near-Solar values in galaxies with velocity dispersions $\sigma \lesssim 100$ km s$^{-1}$ to $\approx 0.1-0.3$ dex super-Solar at velocity dispersions $\approx 300$ km s$^{-1}$. If we interpret this variation in terms of star formation timescales, this implies that the highest velocity-dispersion galaxies must have formed over timescales $\lesssim 0.5$ Gyr, though there is considerable scatter depending on which abundance ratio one uses.

Our simulations suggest another possible contributor: the same pattern could be produced if $\epsilon_\mathrm{II/Ia}$ were to increase systematically with velocity dispersion by $\approx 0.3$ dex. None of our simulated cases, which are modelled on the conditions in local-star-forming galaxies, directly probe the much higher gas and star formation densities relevant to ETGs forming at high redshift. Thus we cannot yet predict what trends in $\epsilon_\mathrm{II/Ia}$ one expects in such systems. However, our finding that $\epsilon_\mathrm{II/Ia}$ variations at the $\approx 0.3$ dex level seen in the data are easily produced even over the limited range of galaxy properties sampled by nearby star-forming galaxies suggests that one should strongly consider the possibility of preferential loss of type Ia ejecta as a contributing factor to the observed abundance ratios in ETGs. Indeed, doing so might somewhat relax the need for ETGs to form over such incredibly short timescales.

One can make similar observations about the reliability of IMF diagnostics from abundance ratios. Observed variations in abundance ratios that have been taken as evidence of IMF variation are typically at the level of a few tenths of a dex (e.g., see Figures 2 - 5 of \citealt{Venn+04}), well within the range of variation that we have shown differential metal loading to be capable of creating. The largest claimed observed abundance ratio variations are for $s$-process to iron peak elements (e.g., the Ba/Fe ratio -- Figure 1 of \citealt{Tsujimoto11a}), and these can reach almost 1 dex, which is too large to be explained entirely differential metal loading at the levels we find in our simulations, but even in this case differential metal loading could make a non-negligible contribution.

On the theoretical side, \citet{Recchi+14} predict that in their ``IGIMF'' model for how the IMF depends on galaxy star formation rates, dwarf galaxies with star formation rates $\lesssim 10^{-2}$ M$_\odot$ yr$^{-1}$ could show $\alpha$/Fe ratios up to $\approx 0.7$ dex smaller than stars in the Milky Way at similar Fe/H, though the variation is much smaller if they also assume that the IMF depends on galaxy metallicity. The larger prediction for the case of a metallicity-independent but SFR-dependent IMF would yield a variation large enough not to be confused by differential metal loading effects, but the latter would not -- and \citet{Lacchin+20} show that nearby ultra-faint dwarf galaxies have $\alpha$/Fe ratios that are much close to the Milky Way value than the predictions at the large end of the IGIMF prediction range, which would seem to rule out the possibility of an effect large enough to be safe from confusion with differential metal loading.

In summary, the possibility of differential metal loading at the level we have measured in our simulations appears to call into question at least some of the inferences about the star formation process that previous authors have made based on elemental abundance ratios. It is clearly an urgent task to clarify the dependence of the differential metal loading on galaxy properties so that we can characterise not just the typical size of the variations it induces, but their expected direction.

\subsection{Caveats}

No discussion of numerical simulations is complete without a recitation of their limitations and approximations. The general limitations of the QED simulation suite -- that supernova feedback is not yet treated self-consistently, that the simulations do not yet include magnetic fields, and that the geometry of the ``tall box'' means that the results are unavoidably-dependent on exactly how one treats the boundary conditions at the vertical faces of the simulation domain -- have been discussed extensively in \citetalias{QEDI} and \citetalias{QEDIII}, and we will not repeat them here. Instead, we will focus on caveats that are particular to the current set of simulations and the conclusions we draw from them regarding differential metal loading.

One important caveat is that our values of $\epsilon$, the differential retention factor, are measured within a few kpc of the galactic plane, and thus do not account for fountain flows that fall back after reaching larger distances. It is conceivable that, in a larger volume, mass return might reduce the differential between elements from different nucleosynthetic sources. That is, we have found that in \fid, to take one example, type Ia ejecta are lost at higher rates than type II ejecta -- but it is possible that this simply means that type Ia ejecta manage to reach typical heights of 5 kpc before falling back, whereas type II ejecta typically fall back after reaching only $\sim 2$ kpc from the plane, but that once fallback is taken into account the fraction of type Ia and type II ejecta that reach CGM distances of tens of kpc are similar. Such an outcome appears improbably given that we find that differential loading is established very close to the disc plane, and that the outflowing material is very well-mixed by the time it reaches the edge of the simulation domain, but given the limitations of the tall box geometry we cannot rule it out entirely.

A second important caveat is that our results are dependent on our assumptions about the relative scale heights of different nucleosynthetic sources, which we have taken from observations. Those observations, however, are clearly limited. Our AGB scale heights come from observations of the Milky Way \citep{Jackson02a}, and it is unknown how the ratio of AGB scale height to the scale heights of type Ia or type II supernovae vary from one galaxy to another. Similarly, our ratio of type Ia to type II scale heights is derived from a simple of only $\approx 100$ SNe in relatively nearby spiral galaxies \citep{Hakobyan17a}; it is unclear how the results generalise to either dwarf galaxies or to conditions beyond the local Universe.

Ideally one would determine the relative scale heights of the different nucleosynthetic sources from a self-consistent simulation, though this present a formidable computational challenge: because the delay time distributions for element return by type Ia SNe and AGB stars are comparable to galactic gas depletion times, a self-consistent simulation likely needs to be fully cosmological so as to include ongoing gas accretion to replace the gas consumed by star formation or ejected in winds while the simulation runs. However, this is challenging to combine with the extremely high resolution requirements -- $\sim\mathrm{few}$ pc resolution not just at the highest densities, but throughout the kpc-sized region around it where the wind phase structure is established -- that we showed in \citetalias{QEDI} are required to obtain converged results for wind metal loading. A useful future strategy thus might be fully cosmological simulations to establish the properties of the different stellar populations responsible for different nucleosynthetic channels, followed by zoom-in simulations that achieve much higher resolution for shorter run times that establish the resulting wind metal loadings.

\section{Conclusions}
\label{sec:conclusion}

In this paper we use a suite of high-resolution simulations of galaxies covering a range of gas surface densities and metallicities to explore the extent to which newly-synthesized metals from different stellar sources -- type II supernovae, type Ia supernova, and AGB stars -- are promptly lost to galactic winds. Each of these sources differs in vertical distribution relative to the galactic disc and in the specific thermal energy with which the newly-synthesized elements are returned, and thus it is plausible that different proportions of them might be lost, which would in turn have important implications for models that seek to infer galaxy properties such star formation timescales or initial mass functions from abundance ratios.

Our simulations show that differential metal loading does in fact occur, with the fraction of a given element retained in the ISM rather than being lost to the wind typically varying by $\sim 0.3$ dex from one element to another. There is, however, no simple rule about which elements will suffer greater or larger losses -- our suite includes examples where each of the element types is preferentially lost. Which one element suffers preferential loss appears to be the result of a complicated interaction between the spatial distributions of the injection sites and the interstellar medium, and for the two supernovae sources there is a good but not perfect correlation between fraction of newly injected elements lost and a simple estimate of its volume filling fraction of supernova remnants based on the combination of supernova rate and vertical density and temperature distribution.

Our finding that differential metal loading of galactic winds can induce variations at the $\approx 0.3$ dex level in the rates at which galaxies are enriched by different nucleosynthetic sources has important implications for the interpretation of abundance ratio diagnostics. The signals upon which these diagnostics work are often at similar levels, and thus there is significant potential for differential metal loading to masquerade as another process. As an example, systematic variations in the $\alpha$ to Fe ratios of early type galaxies with galaxy velocity dispersion have been taken as evidence for very rapid star formation in high velocity dispersion systems, but we show here that the same trend could be produced if low velocity dispersion systems suffer little differential metal loss, while high-velocity dispersion systems preferentially lose type Ia elements at rates comparable to those in some of the simulations in our suite.

Our work here is limited by our simulation geometry, which is a ``tall box'' that prevents us from including more compact and starbursting systems, and makes it impossible for us to follow fallback of ejecta that reach distances more than a few kpc from the disc. Thus while our simulations show that some types of metals are definitely lost more than others, they do not yet resolve the question of whether those lost metals are eventually re-accreted, and on what timescales. In future work we intend to carry out similar simulations that move beyond the tall box geometry, and will give us a fuller picture of galactic wind metal loss over a larger volume.

\section*{Acknowledgements}

AV and MRK acknowledge support from the Australian Research Council through its Laureate Fellowship scheme, award FL220100020. This research was undertaken with the assistance of resources (award jh2) from the Pawsey Supercomputing Research Centre’s Setonix Supercomputer (\url{https://doi.org/10.48569/18sb-8s43}), with funding from the Australian Government and the Government of Western Australia, and from the National Computational Infrastructure (NCI Australia), an NCRIS enabled capability supported by the Australian Government.

\section*{Data Availability}

The data used in this paper can be shared upon reasonable request to the corresponding author.



\bibliographystyle{mnras}
\bibliography{references} 




\appendix




\bsp	
\label{lastpage}
\end{document}